\documentclass[aps,twocolumn,superscriptaddress,nofootinbib,prb]{revtex4-2}

\usepackage{graphicx}
\usepackage{placeins} 
\usepackage{amssymb,amsmath,amsfonts,bm}
\usepackage[normalem]{ulem}
\usepackage{hyperref}
\usepackage[dvipsnames]{xcolor}
\hypersetup{urlcolor=RoyalBlue, colorlinks=true,citecolor=RoyalBlue,linkcolor=RoyalBlue} 
\usepackage{physics}
\usepackage[final]{showkeys}
\usepackage{float}

\begin{document}
\title{$Z_3$ confined and deconfined Coulomb liquids in $S_{\rm eff} = 3/2$ pyrochlore magnets}

\author{Jay Pandey}
\affiliation{Department of Theoretical Physics, Tata Institute of Fundamental Research, Mumbai 400005}
\author{Souvik Kundu}%
\affiliation{International Center for Theoretical Sciences, Tata Institute of Fundamental Research, Bengaluru 560089}
\author{Kedar Damle}
\affiliation{Department of Theoretical Physics, Tata Institute of Fundamental Research, Mumbai 400005}

\date{\today}

\begin{abstract}
We identify an interesting regime in the physics of pyrochlore magnets in which spin-orbit and crystal field effects lead to {\em two} low-lying magnetic doublets that can be modeled as an effective spin $S=3/2$ degree of freedom that sees a dominant easy-axis antiferromagnetic exchange $J>0$ favoring the local $[111]$ axes, which competes with a comparably strong single-ion anisotropy $\Delta = J+\mu/2$ (with $|\mu| \ll J$) favoring the perpendicular planes. For a precise analysis, we study the $T/J \rightarrow 0$ limit in which $w \equiv \exp(-\mu/T)$ is the control variable. 
In this limit, we find {\em two topologically distinct} zero-field Coulomb phases separated by a first-order $Z_3$ confinement transition at $w_c \approx 2.02$. Both Coulomb phases admit a description in terms of the fluctuations of a coarse-grained divergence-free polarization field. However, the flux of this polarization field is restricted to integer multiples of $3$, and only charges that are multiples of 3 are deconfined in one of these phases, while all integer fluxes are allowed and all integer charges are deconfined in the other phase. Experimental systems with small negative $\mu$ ({\em i.e.}, $-J \ll \mu < 0$) are therefore predicted to exhibit signatures of this topological transition when cooled below $T_c \approx 1.42|\mu|$.
\end{abstract}

\maketitle

\section{Introduction}
The ground states of some strongly correlated condensed matter systems and their low-lying excitations are best understood in terms of new fractionalized degrees of freedom and emergent gauge structures~\cite{Wen_QFTofMBS_book,Wen_RevModPhys_2017,Wen_Science_2019,Simon_TopoQuant_book}; fractional quantum Hall systems provide prototypical examples of this behavior~\cite{GirvinYang_book,Sachdev_QuantumPhasesofMatter_book,MooreMoessner_book}. The possibility of having such physics in insulating magnets with competing antiferromagnetic interactions $J$ has also attracted considerable attention~\cite{Mila_Mendels_Lacroix_2011book,MooreMoessner_book}. Although many promising candidates exist~\cite{Broholm_Cava_Kivelson_Nocera_Norman_Senthil_Science_2020,
Clark_Abdeldaim_Annual_Review_of_Material_Research_2021}, an unequivocal identification of experimental signatures of such fractionalized degrees of freedom has remained somewhat controversial in such magnets. Fortunately, such spin systems can also have effectively classical low-temperature regimes in which the long-wavelength spin correlations feature unmistakable signatures of similar emergent degrees of freedom. 

Insulating rare-earth oxides in which the magnetic ions occupy sites of a pyrochlore lattice (comprising a regular arrangement of corner-sharing tetrahedra) provide a well-studied class of examples of such behavior~\cite{Harris_etal_1997,Siddharthan_etal_1999,Ramirez_etal_1999,Bramwell_etall_2001, Bramwell_Gingras_review_2001_science, Fennel_etal_2004, Castelnovo_Moessner_Sondhi_2008,  Fennel_etalScience2009,Henley_Coulombphasesreview2010,Castelnovo_Moessner_Sondhi_review2012, Bramwell_Harris_review2020}. At temperatures $T \ll J$, such ``classical spin-ice'' materials display Coulomb spin liquid behavior: Their long-wavelength spin correlations are given by the correlations of a fluctuating divergence-free polarization field $\vec{P} = \nabla \times \vec{a}$, where $a$ is an emergent U($1$) gauge field. 
In well-studied examples of this behavior ({\em e.g.}, Ho$_2$Ti$_2$O$_7$ and Dy$_2$Ti$_2$O$_7$)~\cite{Bramwell_etall_2001, Bramwell_Gingras_review_2001_science, Fennel_etal_2004, Castelnovo_Moessner_Sondhi_2008,  Fennel_etalScience2009,Henley_Coulombphasesreview2010,Castelnovo_Moessner_Sondhi_review2012, Bramwell_Harris_review2020}, 
crystal-field effects isolate a single low-energy doublet $F^z=\pm F$ of single-ion states from the spin-orbit-coupled ground state multiplet of total angular momentum $F$ ({\em e.g.}, $F=8$ for Ho$_2$Ti$_2$O$_7$). Here $F^z$ is the component along the local $[111]$ (tetrahedral body diagonal) axis. The low-energy physics is thus of {\em effective $S=1/2$ variables}~\cite{Rosenkaraz_etal_2000, Bramwell_Gingras_review_2001_science, Harris_etal_1997} $\vec{S}_r$ which see a frustrated Ising interaction $J$ that couples only these $z$ components. 

Here, we predict the existence of new phenomena featuring an emergent $Z_3$ gauge structure in addition to this U(1) gauge field for a class of such pyrochlore magnets. We expect this prediction to hold whenever crystal-field and spin-orbit effects lead to not one well-isolated doublet, but {\em two} low-energy doublets that can be modeled as an effective $S=3/2$ degree of freedom that sees a single-ion anisotropy $\Delta$ that favors the plane perpendicular to the local $[111]$ axes and thereby {\em competes} with the easy-axis Ising exchange $J$ that favors these local axes.
Specifically, we demonstrate that the competition between these two scales leads to {\em two topologically distinct} low-temperature Coulomb phases separated by a $Z_3$ confinement transition, with both phases featuring Coulomb liquid behavior in exactly the same sense as in Ho$_2$Ti$_2$O$_7$ and Dy$_2$Ti$_2$O$_7$.

\begin{figure}
    \includegraphics[width=\linewidth]{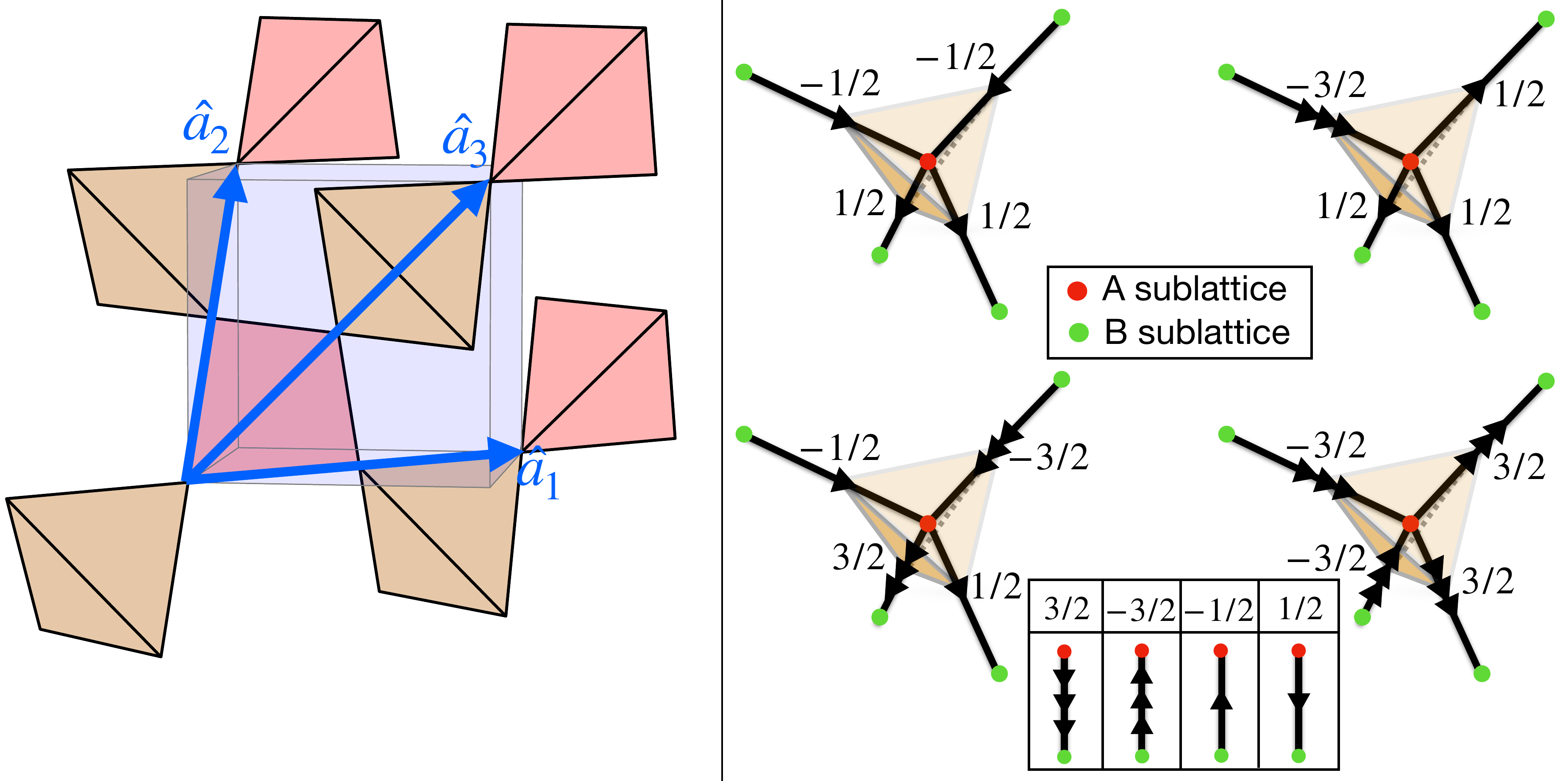}
    \caption{Left: Bravais lattice translations $\vec{a}_{1,2,3}$ of the pyrochlore lattice whose sites lie at the centers of the bonds of a diamond lattice. Right: The mapping from spin $S^z$ variables to the lattice polarization field $\vec{P}$ on bonds of the diamond lattice.}
    \label{fig:ThreeHalfSchematics}
\end{figure}

\begin{figure*}
  (a)\includegraphics[width=0.3\linewidth]{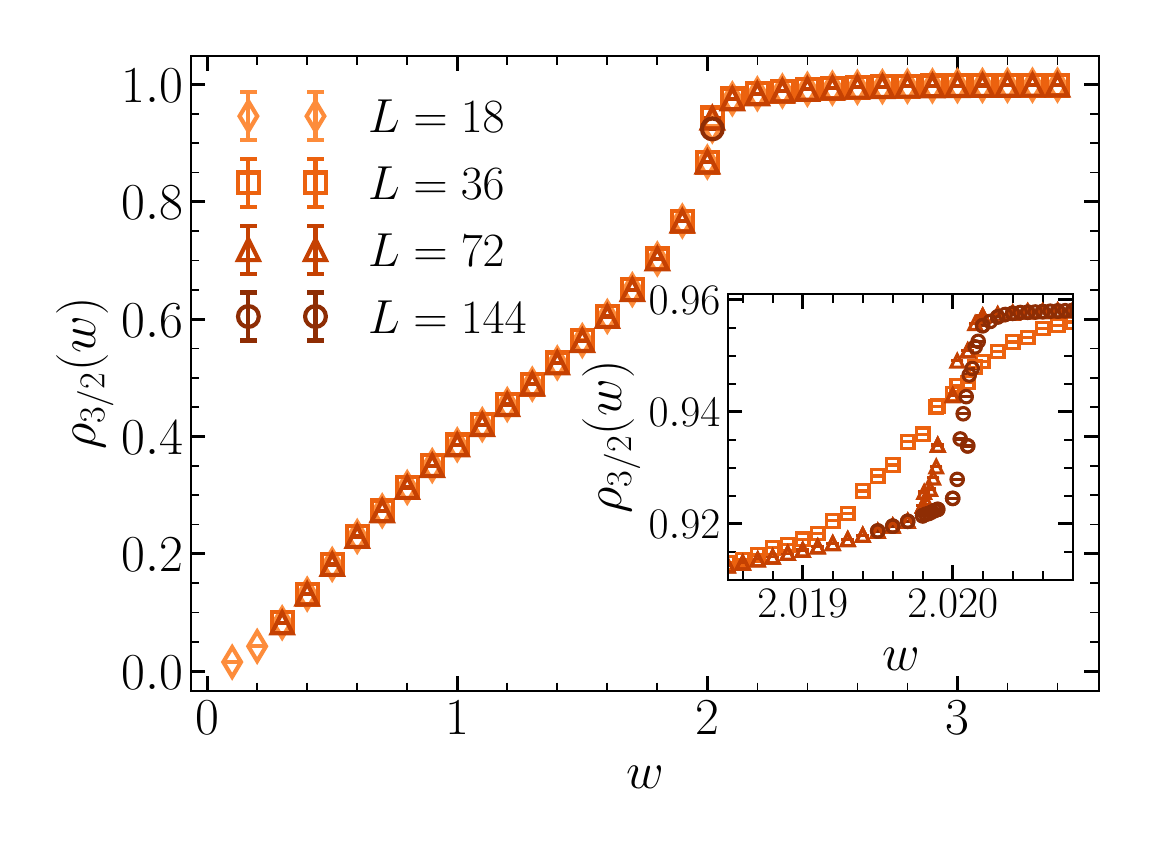}
     (b)\includegraphics[width=0.3\linewidth]{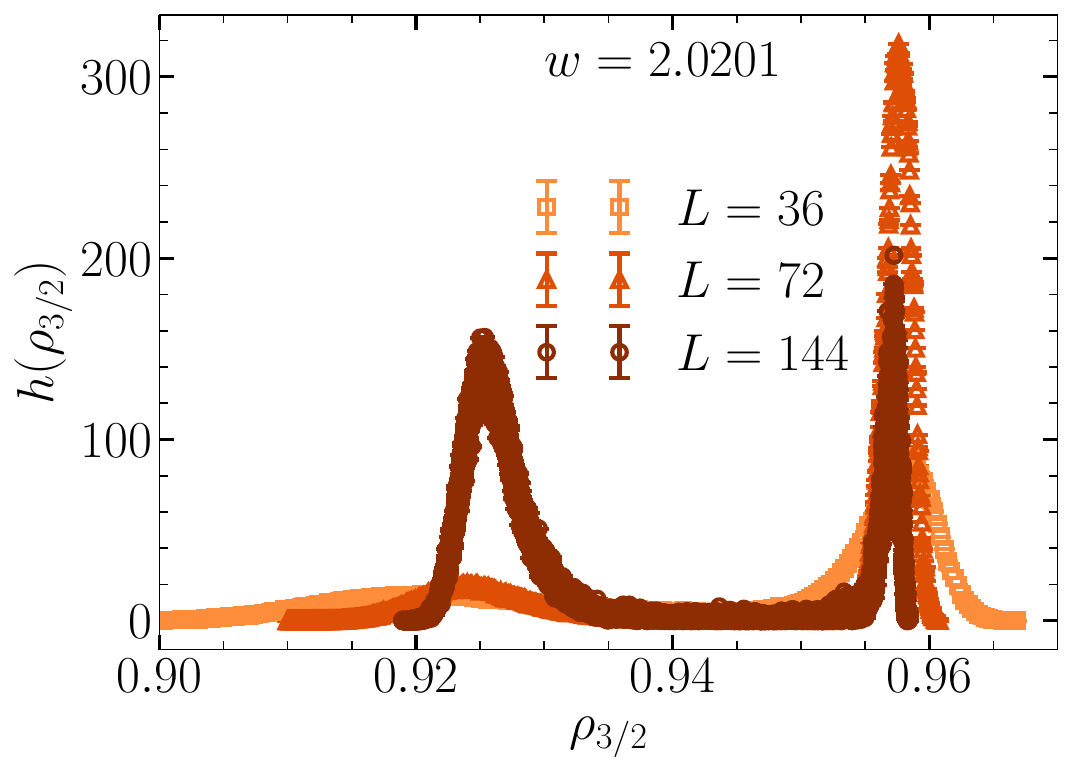}
        (c)\includegraphics[width=0.3\linewidth]{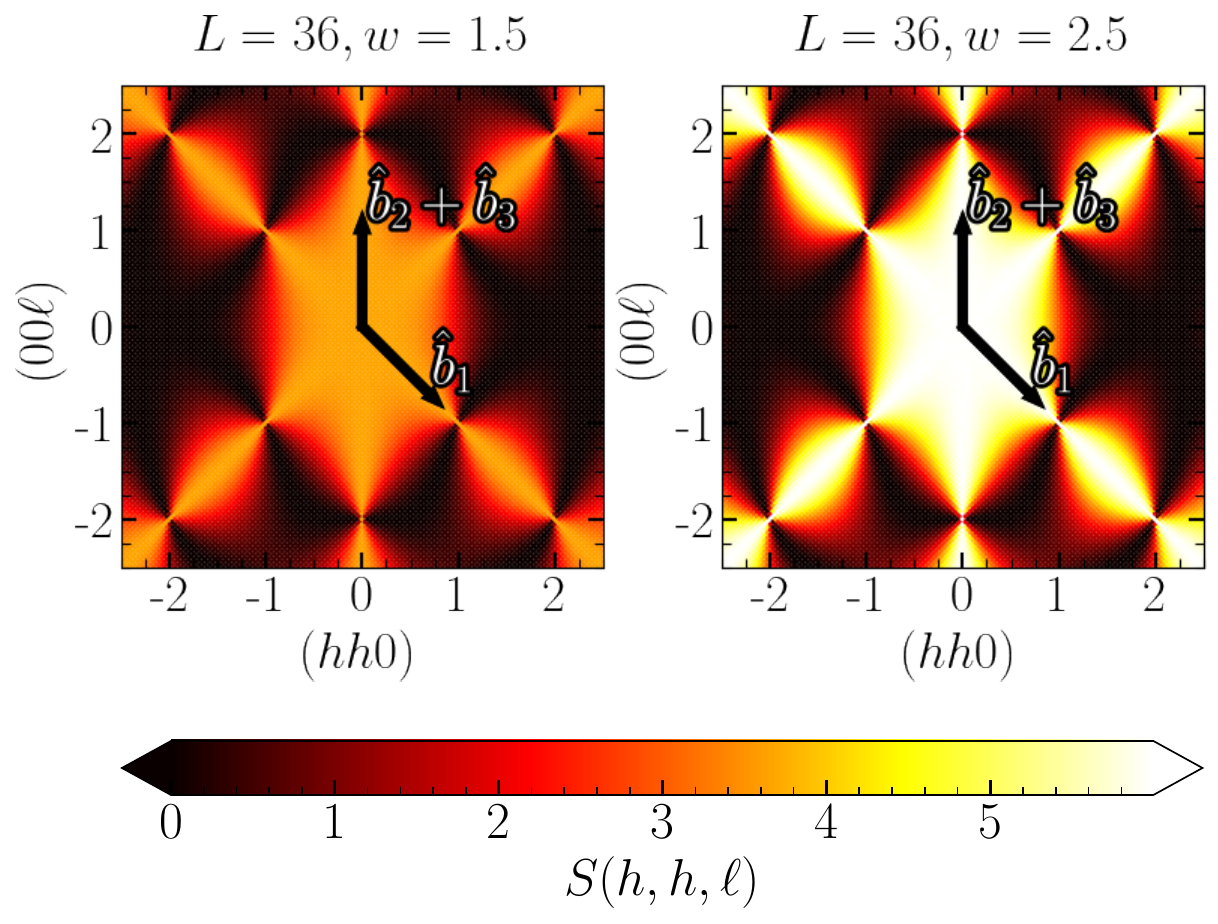}
        \caption{(a) $\rho_{3/2} = \epsilon/\mu $, the energy density in units of $\mu$, has a clear jump at $w\approx 2.02$, corresponding to a first-order transition with a latent heat.
        (b) Its histogram shows a clear two-peak structure in the vicinity of this transition. (c) The spin-flip structure factor in both phases has pinch-point singularities characteristic of a Coulomb liquid.} 
        \label{fig:Rho3by2Jump}
\end{figure*}

Although the spin structure factor in {\em both} phases exhibits pinch-point singularities that can be understood in terms of fluctuations of $P$~\cite{Henley_Coulombphasesreview2010,Huse_Krauth_Moessner_Sondhi_2003}, they are nevertheless separated by a first-order thermodynamic phase transition. This transition is a $Z_3$ confinement transition: When periodic boundary conditions are imposed, the flux of the divergence-free polarization field is restricted to integer multiples of three in the $Z_3$ confined Coulomb phase, while all integer-valued flux sectors contribute to the partition function in the other Coulomb phase. Equivalently and independent of boundary conditions, all integer-valued test charges (that serve as sources for the polarization field) are deconfined in the $Z_3$ deconfined Coulomb phase, while only test charges which are multiples of $3$ remain deconfined in the $Z_3$ confined Coulomb phase. A magnet with $\Delta \lesssim J$ can be driven from the $Z_3$-deconfined to the $Z_3$ confined Coulomb phase simply by going to low-enough temperatures, while systems with $\Delta \gtrsim J$ remain in the flux-deconfined Coulomb phase down to the lowest temperatures.

\section{Effective model and preliminary analysis}
In the rest of our discussion, we focus on the following effective Hamiltonian that incorporates the essentials of this physics: 
\begin{eqnarray}
    H=  \sum_{\langle r, r' \rangle} \left( J_{\parallel} S_r^z S_{r'}^z + J_{xy} (S_r^x S_{r'}^x + S_r^y S_{r'}^y) \right) + \Delta \sum_r (S_r^z)^2 \;, \nonumber \\
    &&
\end{eqnarray}
where $\vec{S}_r$ are effective spin $S=3/2$ degrees of freedom on pyrochlore sites, $J_{\parallel} > 0$ is the dominant Ising-like nearest-neighbor coupling along the local $[111]$ axes, $\Delta > 0$ is a comparably large single-ion anisotropy which has an easy-plane character that favors spin orientations perpendicular to the local $[111]$ axis, the transverse coupling $J_{xy}$ is assumed to be small compared to both $J_{\parallel}$ and $\Delta$ as well as their difference, and we have left out additional subdominant terms that are no larger in energy scale than $J_{xy}$.

To encode this hierarchy of scales, we write $J_{\parallel} = J>0$ and $\Delta = J+\mu/2$, with the understanding that $J_{xy} \ll |\mu| \ll J$.
At low temperatures in the range $J_{xy} \ll T \ll J$, quantum fluctuations are negligible and configurations that do not have the minimum possible Ising-exchange energy have exponentially small Boltzmann weight. The latter observation implies that the physics in this regime of interest is controlled by thermal fluctuations {\em within the set of minimally-frustrated configurations that minimize the easy-axis exchange energy}. Our key observation is that the character of these fluctuations is controlled by the ratio of $\mu/T$ on which we can place no strong restrictions in the regime of interest, except that its sign is a fixed microscopic property of each system, and its magnitude cannot be too small ({\em i.e.}, $T$ cannot be very much larger than $\mu$). 
In particular, $|\mu|/T$ can take on very large values if the system is cooled sufficiently below the scale set by $|\mu|$. 
Here, we parameterize this variation by $w=\exp(-\mu/T)$, which controls the relative Boltzmann weights of the minimally frustrated configurations. Of course, a given system will only explore about half of the $w$ axis, but this still leaves open the possibility of interesting low-temperature behavior in particular cases, as we demonstrate by delineating the phase diagram as a function of $w$.

For a precise analysis of this physics, we focus on the limit $T/J \rightarrow 0$ with $w$ held fixed and $J_{xy} = 0$, in which the classical partition function only gets contributions from these minimally frustrated configurations.
It is convenient to represent these by configurations of a lattice polarization field $\vec{P}_{AB}$ defined on bonds of the diamond lattice whose $A$ ($B$) sublattice sites represent the body centers of the ``up-pointing" (``down-pointing'') pyrochlore tetrahedra and bonds host the pyrochlore spins $S^z_r$ at their centers (Fig.~\ref{fig:ThreeHalfSchematics}). Thus we write $\vec{P}_{AB} = S^z_r\hat{e}_{AB}$, where $\hat{e}_{AB}$ is oriented from the $A$ sublattice site to the $B$ sublattice site connected by this bond. 

All the minimally frustrated configurations that contribute to the partition function are characterized by a constraint that the total spin $S^z_{\rm tet}$ on each tetrahedron must be zero: $S^z_{\rm tetra} = 0$. In terms of the lattice polarization field, this implies that its lattice divergence is zero: $\Delta \cdot \vec{P} = 0$. Such configurations can locally satisfy this constraint in four ways: All four spins have $|S^z_r| = 1/2$. Or three spins have $|S^z_r| = 1/2$, and one has $|S^z_r| = 3/2$. Or, two spins have $|S^z_r| = 1/2$, and the other two have $|S^z_r| = 3/2$. Or all four have $|S^z_r| = 3/2$. 
The utility of the parameterization in terms of $w$ is immediately clear when one notes that the relative Boltzmann weight of minimally frustrated configurations has one factor of $w$ for each $S^z_r = \pm 3/2$, while each $S^z_r = \pm 1/2$ contributes a factor of unity to the weight of a configuration. Thus, the low-energy properties are to be computed from the following partition function:
\begin{equation}
Z = \sum_{C} w^{n_{3/2}(C)} \; ,
\end{equation}
where $n_{3/2}(C)$ is the number of spins with $|S^z_r| = 3/2$ in a minimally frustrated configuration $C$. Thus, the low energy theory is a $44$-vertex model~\cite{Cardy_2001,Baxter_1989} on the diamond lattice.
\begin{figure*}
    \includegraphics[width=0.3\linewidth]{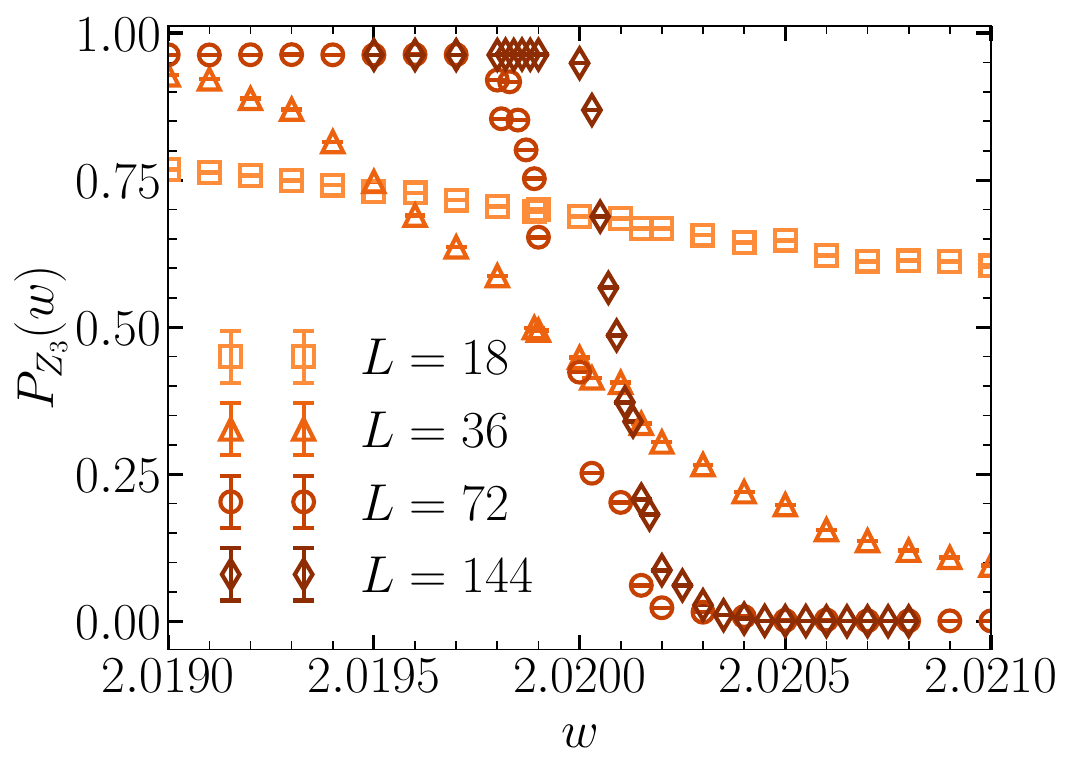}
    \includegraphics[width=0.3\linewidth]{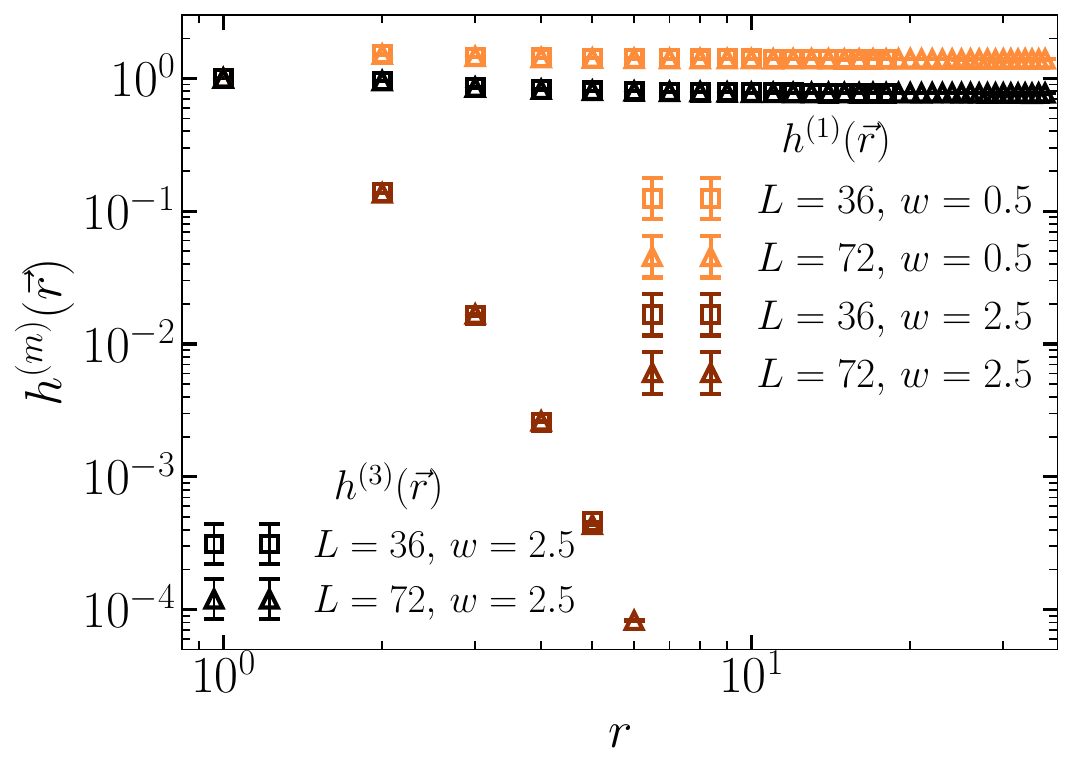}
    \includegraphics[width=0.3\linewidth]{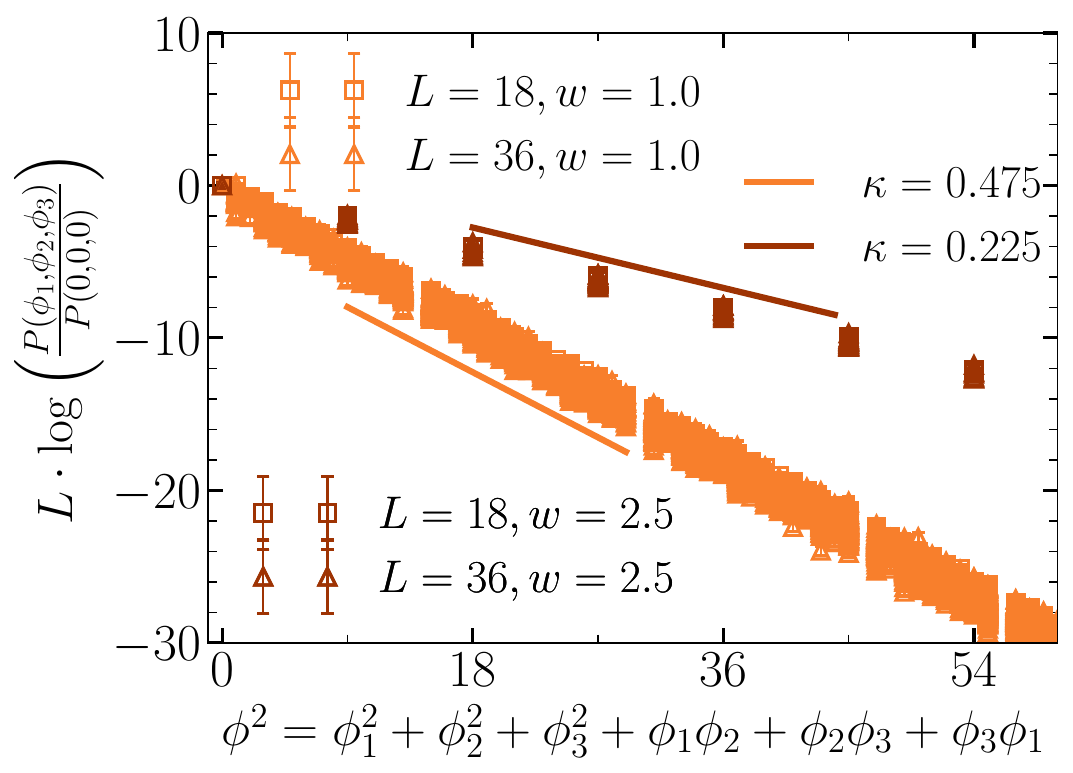}
    \caption{(a) The probability $P_{Z_3}$ of having a nonzero $Z_3$ flux, {\em i.e.}, of the flux vector $\vec{\phi}$ having a component not divisible by $3$, tends to zero (remains nonzero) at large $L$ for $w>w_c$ ($w<w_c$). (b) The histogram $h^{(1)}(\vec{r})$ of head-to-tail distances of unit-charge worms goes to a nonzero constant at large $r$ for $w<w_c$,  but decays rapidly to zero for $w>w_c$, while the corresponding $h^{(3)}(\vec{r})$ for charge $3$ worms goes to a nonzero constant at large $r$ for $w>w_c$. (c) The probability distribution $P(\vec{\phi})$ of $\vec{\phi}$ is a Gaussian for both $w>w_c$ and $w<w_c$, but only $\vec{\phi}$ with all components divisible by $3$ survive at large $L$ and follow the Gaussian for $w>w_c$.}
    \label{fig:Pfrac}
\end{figure*}

In classical spin-ice materials with pseudo-spin-$1/2$ moments, the existence of a lattice-level divergence-free polarization field is a crucial ingredient of the theoretical description in terms of a coarse-grained version of such a polarization field, from which flow predictions for Coulomb spin liquid behavior and its signatures, namely, dipolar pinch-point singularities in the structure factor~\cite{Henley_Coulombphasesreview2010,Castelnovo_Moessner_Sondhi_review2012,
Huse_Krauth_Moessner_Sondhi_2003}. 
From the vantage point of this theory, it seems reasonable to expect that any information about the precise values allowed for the microscopic lattice-level polarization field would play no significant role in the coarse-grained effective theory.
In other words, one might expect the physics at very small $w \ll 1$ to be continuously connected to that at $w \gg 1$, and such pseudospin $S=3/2$ magnets to also have essentially the same Coulomb liquid behavior as the pseudospin $S=1/2$ classical spin ice materials. Our key result is that this reasonable-sounding argument turns out to be misleading: In fact, there are two distinct equilibrium Coulomb liquid phases as a function of $w$, separated from each other by a thermodynamic phase transition that has a first-order character, and corresponds to a $Z_3$ flux confinement phenomenon. 

\section{Computational Results}
We have mapped out the equilibrium phase diagram as a function of $w$ using classical Monte Carlo simulations of pyrochlore samples with $L^3$ unit cells and periodic boundary conditions in the three principal directions (Fig.~\ref{fig:ThreeHalfSchematics}). We use a worm algorithm that is modeled on earlier applications of such algorithms to problems with local constraints~\cite{Kundu_Damle_preprint2025,Kundu_thesis2024,Kundu_Damle_PRX_2025,Morita_Lee_Damle_Kawashima2023,
Rakala_Damle_2017,Desai_Pujari_Damle_2021,Alet_etal_PRE_2006,Sandvik_Moessner_2006,Alet_Sorensen_2003,Alet_Sorensen2003B}. In outline, this proceeds as follows: By changing $S^z_r$ on two sites of a down-pointing (up-pointing) tetrahedron without violating the constraint on it, one places two oppositely-charged defects on the two neighboring up-pointing (down-pointing) tetrahedra, which now have $S^z_{\rm tet} = \pm 1$ and host the head and the tail of the worm. The worm update consists of executing a guided random walk for the head with the displacement probabilities obeying local detailed balance, while keeping the tail fixed. When the head reunites with the tail, one has a new minimally-frustrated configuration which can be accepted with unit probability. The histogram $h^{(1)}(\vec{r})$ of head-to-tail distances $\vec{r}$ of unit-charge worms also serves as the estimator for the equilibrium correlation function $C^{(1)}(\vec{r})$ of a pair of $\pm 1$ test charges. The corresponding histograms $h^{(m)}(\vec{r})$ obtained from charge-$m$ worms ($m=2,3$) provide partial information about the correlator $C^{(m)}(\vec{r})$ of two test $\pm m$ charges, in the sense that deconfinement of $h^{(m)}$ implies the same for $C^{(m)}$, but short-ranged behavior of $h^{(m)}$ does not unequivocally imply confinement for $C^{(m)}$~\cite{Kundu_Damle_preprint2025,Kundu_thesis2024}. 

The energy density $\epsilon$ of the system is proportional to the density $\rho_{3/2} = n_{3/2}/L^3$ of $S^z_r = \pm 3/2$ spins: $\epsilon = \mu \rho_{3/2}$. 
From our computations of the $w$ dependence of $\rho_{3/2}$, we note that $\rho_{3/2}$ shows a clear $L$-independent jump at $w_c \approx 2.02$ (Fig.~\ref{fig:Rho3by2Jump}(a)). This jump signals a first-order phase transition with a sizeable latent heat. Thus, we provisionally conclude that there are two distinct thermodynamic phases separated by this transition along the $w$ axis. This conclusion is confirmed by a study of the histograms of $\rho_{3/2}$ at $w_c$ and in its very close vicinity. These show a clear $L$-independent two-peak structure that signals exactly the kind of phase coexistence one expects at a first-order phase transition (Fig.~\ref{fig:Rho3by2Jump}(b)). At a microscopic level, this first-order transition is in fact associated with the percolation of clusters of spins with $S^z= \pm 1/2$ (Appendix).

What distinguishes these two phases at a macroscopic level? To try and answer this, we first examine the experimentally measurable spin-flip structure factor~\cite{Chung_Goh_Mukherjee_Jin_Lozano_Gingras__PRL2022} and ask if there is any qualitative difference between the two phases. From Fig.~\ref{fig:Rho3by2Jump}(c), we see that there is no qualitative difference in the dipolar character of the structure factor and the corresponding pinch-point singularity visible in it. The only change is an overall reduction in intensity in the small-$w$ phase, arising from a larger propensity for the spins to be in the $S^z = \pm 1/2$ doublet compared to the $S^z = \pm 3/2$ doublet. Thus, both phases are Coulomb spin liquids.

To resolve this puzzle, we exploit the periodic boundary conditions of our numerics and note that the total flux of the polarization field in each periodic direction (Fig.~\ref{fig:ThreeHalfSchematics}) for a given configuration is a well-defined (independent of the measurement surface) three-vector $\vec{\phi} \equiv (\phi_1,\phi_2,\phi_3)$ of integers. Moreover, $\vec{\phi}$ cannot be changed by any purely local changes to the configuration. To see why, we note that the flux in some direction changes by $\pm 1$ if the state of a single spin is flipped from $S^z = +1/2$ to  $S^z = -1/2$ to create a pair of $\pm 1$-charged defects, and one of these charges subsequently winds around the torus before annihilating the other charge. Clearly, this process induces a global change to the configuration. Also, this is the smallest change possible in any component of the flux vector $\phi$. Thus, the fluxes are {\em integer-valued topological properties} of a configuration.

Next, we note that the same observation also implies that these integers would have necessarily been restricted to multiples of $3$ if the spins could only take on values $S^z = \pm 3/2$. Motivated by this, we study $w$ dependence of the equilibrium probability $P_{Z_3}$ of having a nonzero $Z_3$ flux, {\em i.e.}, the probability that some component of $\vec{\phi}$ is not divisible by $3$. From Fig.~\ref{fig:Pfrac} (a), we see that $P_{Z_3}$ goes to zero in the thermodynamic limit for $w >w_c$, but has a nonzero large-size limit for $w < w_c$. Thus, we conclude that there are two topologically distinct Coulomb phases separated by 
a $Z_3$ flux confinement transition. 
Although this topological characterization of the transition relies on periodic boundary conditions, we have checked that the thermodynamic transition itself is independent of boundary conditions.
Indeed, there is another intrinsic (independent of boundary conditions) characterization of the transition: We find that the histogram $h^{(1)}(\vec{r})$ of head-to-tail distances $\vec{r}$ of unit-charge worms, which serves as the estimator for the unit test charge correlator $C^{(1)}(\vec{r})$, is deconfined (goes to a nonzero large-$r$ limit) for $w<w_c$, but is confined (decays rapidly to zero) for $w>w_c$, while the corresponding histogram for charge-$3$ worms shows deconfinement for $w>w_c$, implying the same for $C^{(3)}(\vec{r})$ (Fig.~\ref{fig:Pfrac} (b)). 
With periodic boundary conditions, the probability distribution $P(\vec{\phi})$ of $\vec{\phi}$ is seen in both phases to have the Gaussian form expected for a Coulomb liquid with diamond lattice symmetries (Fig.~\ref{fig:Pfrac}(c))~\cite{Kundu_Damle_preprint2025,Kundu_Damle_PRX_2025,Patil_Dasgupta_Damle_2014}.
The important distinction between $w>w_c$ and $w < w_c$ is that the values of $\vec{\phi}$ that survive the thermodynamic limit and obey this Gaussian distribution are restricted in the large-$w$ phase to those with zero $Z_3$ flux in all directions, whereas all integer sectors fall on this Gaussian in the small-$w$ phase.

To summarize: Unit test charges are deconfined and the fluxes can 
take any integer values for $w<w_c$. When $w>w_c$, unit test charges are confined, but charge-$3$ defects are
 deconfined and fluxes are integer multiples of $3$. This is direct computational evidence that the transition between the two phases is associated with the confinement of an emergent $Z_3$ gauge field. Although this characterization in terms of $Z_3$ confinement is the one that is most directly motivated by our computational results, the classical nature of the problem also permits an alternate but equivalent description in terms of a $Z_3$ spin model: We view the polarization field as the divergence-free current 
   of the current loop representation of a fictitious three-dimensional $xy$-spin model~\cite{Dasgupta_Halperin_1981,Kadanoff_Kirkpatrick_Nelson_1977,Geraedts_Motrunich_2012,Alet_Sorensen_2003,Alet_Sorensen2003B}. This maps $C^{(m)}(\vec{r})$ 
   to the correlator $\langle e^{im\theta(\vec{r})} e^{-im\theta(0)}\rangle$ of the fictitious $xy$ spin 
   $e^{i \theta(\vec{r})}$. In this language, the large-$w$ phase has long-range order for $e^{3i\theta(\vec{r})}$ but short-range order for $e^{i\theta(\vec{r})}$, while $e^{i\theta(\vec{r})}$ is itself long-range ordered in the small-$w$ phase. 
   The transition between these two phases is thus expected to be a $Z_3$ symmetry-breaking transition separating a ferromagnet from the $Z_3$ analog of a nematic phase. Since $Z_3$ spin and gauge models are dual to each other in three dimensions~\cite{Dasgupta_Halperin_1981,Kogut_1979}, this is consistent with the description in terms of a $Z_3$ confinement transition.

\section{Discussion:}
In what sense are there two distinct Coulomb phases and an intervening first-order transition at $T_c = \mu/\log(w_c) \approx 1.42 \mu$? After all, any nonzero temperature $T$ induces a nonzero density of charges, and this charge density is expected in principle to destroy the Coulomb liquid phase in three dimensions. The answer is of course that this density of charges is exponentially small at temperatures $T \ll J$, exactly analogous to the exponentially small charge density in pseudo-spin $S=1/2$ spin-ice materials like Ho$_2$Ti$_2$O$_7$ at low temperature~\cite{Henley_Coulombphasesreview2010,Castelnovo_Moessner_Sondhi_review2012}. 
These well-studied materials nevertheless realize a low-temperature classical Coulomb liquid phase, in the sense that the spin flip structure factor does show clear signatures of the underlying Coulomb physics~\cite{Fennel_etal_2004,Fennel_etalScience2009}.  
In the $S=3/2$ system, we similarly expect the first-order transition at $T_c \approx 1.42|\mu|$ between the $Z_3$-confined and $Z_3$-deconfined liquids to be clearly visible in specific heat measurements for materials with a small negative $\mu$.

Moreover, the distinction between the $Z_3$ confined and deconfined Coulomb phases, being topological in nature, is expected to survive quantum fluctuations induced by the presence of small transverse couplings.
Thus, the ground state phase diagram is expected to feature two topologically distinct quantum Coulomb liquid ground states separated by a transition that maps to the confinement transition of a $3+1$-dimensional quantum $Z_3$ gauge theory. In this specific sense, this $T=0$ physics is expected to realize not just the usual U($1$) gauge structure of quantum Coulomb phases, but also an emergent $3+1$-dimensional $Z_3$ gauge field.

Other recent work~\cite{Kundu_Damle_PRX_2025,Kundu_Damle_preprint2025,Pandey_Damle_Spin1preprint2025} has also identified possible $Z_2$ flux-confinement transitions in anisotropic $S=1$ magnets, underscoring the rich variety of other possibilities that arise as a result of such competing anisotropies. It would therefore be interesting to use microscopic tight-binding descriptions~\cite{Lee_Bhattacharjee_Kim_2013,Rau_etal_2014} for strongly correlated spin-orbit coupled materials to identify a regime with such competing interactions in some materials. In the $S=3/2$ case, our results demonstrate that such materials will indeed have two distinct Coulomb liquid phases in a very real sense, with a transition that can be seen in the specific heat, although the equilibrium spin structure factor will remain qualitatively similar across the transition. Interestingly, parallel work~\cite{Kundu_etal_noneqpreprint2026} finds regimes of slow non-equilibrium quench dynamics for one sign of $\mu$ but not the other, and this could potentially be used as a signal to distinguish between the two topologically distinct equilibrium phases identified here.

\section{Acknowledgements}
We thank Subhro Bhattacharjee, R. Coldea, and H. Takagi for useful discussions. We gratefully acknowledge generous allocation of computing resources by the Department of Theoretical Physics (DTP) of the Tata Institute of Fundamental Research (TIFR), and related technical assistance from K. Ghadiali and A. Salve. The work of JP was supported at the TIFR by a graduate fellowship from DAE, India, while the work of SK at the International Center for Theoretical Sciences of TIFR (ICTS-TIFR)  was supported by a postdoctoral fellowship from DAE, India, under project no. RTI4019.  KD was supported at the TIFR by DAE, India, and in part by a J.C. Bose Fellowship (JCB/2020/000047) of SERB, DST India, and by
	the Infosys-Chandrasekharan Random Geometry Center
	(TIFR).

\bibliography{bibliography}

\clearpage
\section*{Appendix}
\begin{figure*}[b]
a)    \includegraphics[width=0.3\linewidth]{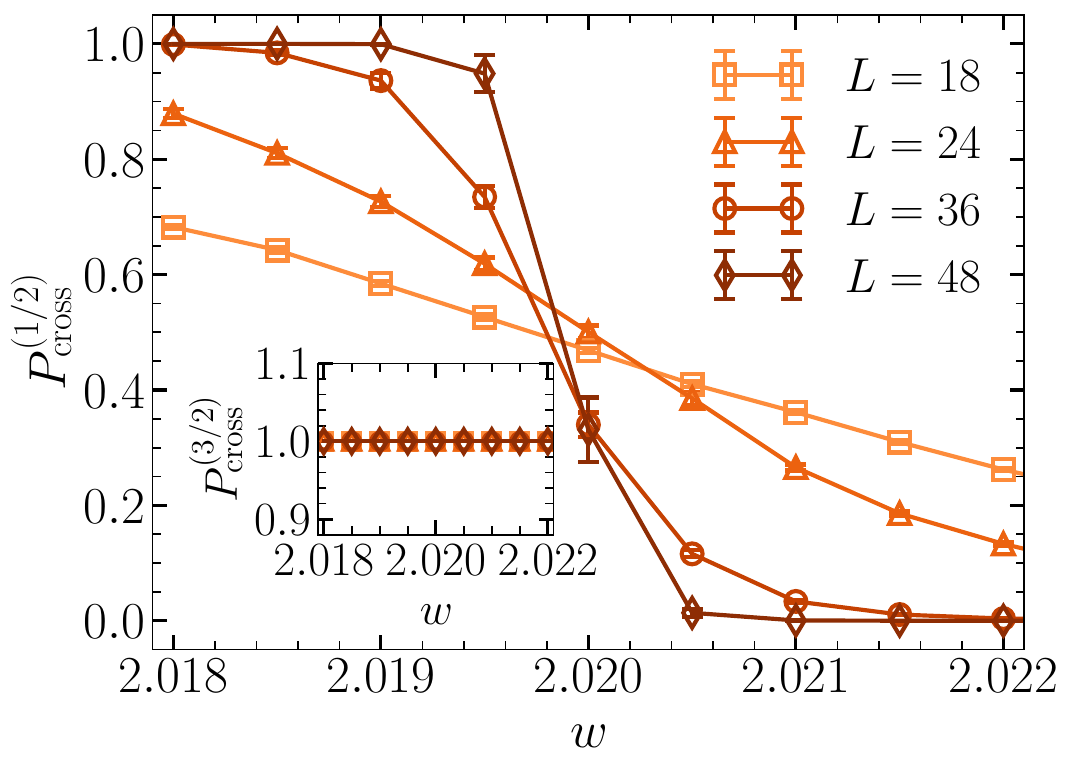}
  b)  \includegraphics[width=0.3\linewidth]{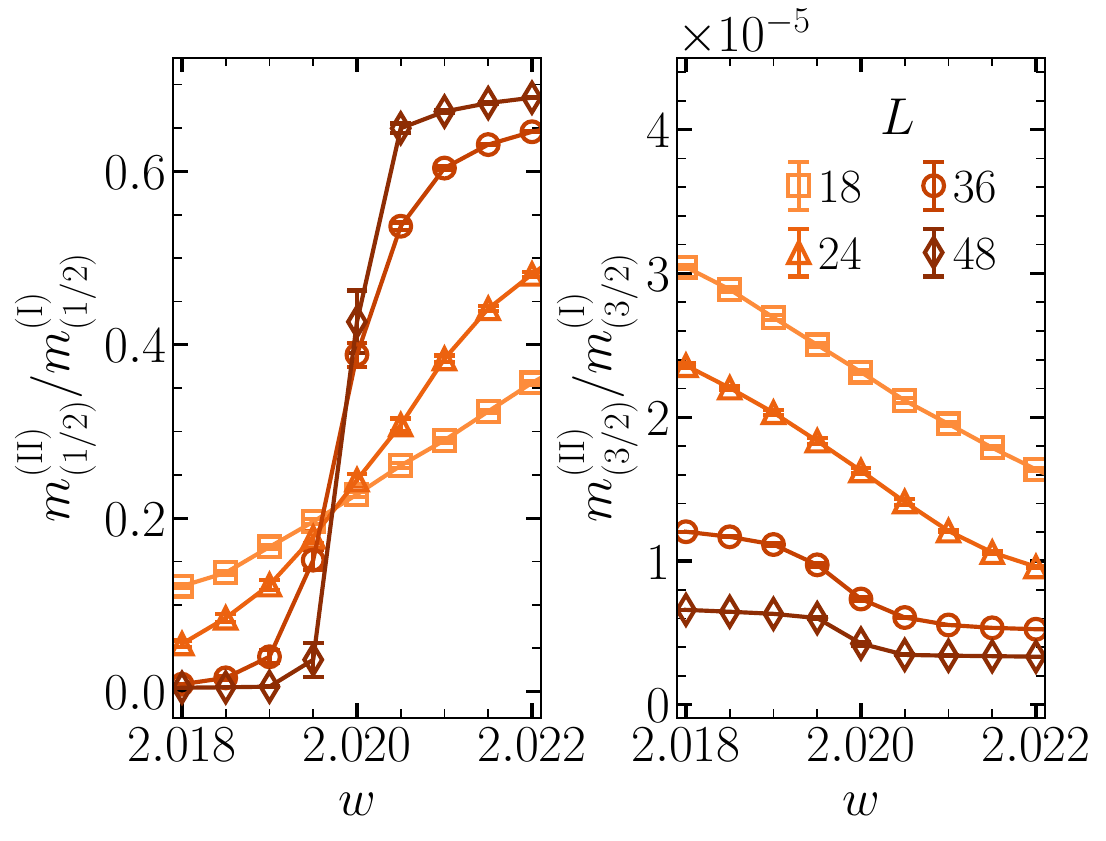}
   c) \includegraphics[width=0.3\linewidth]{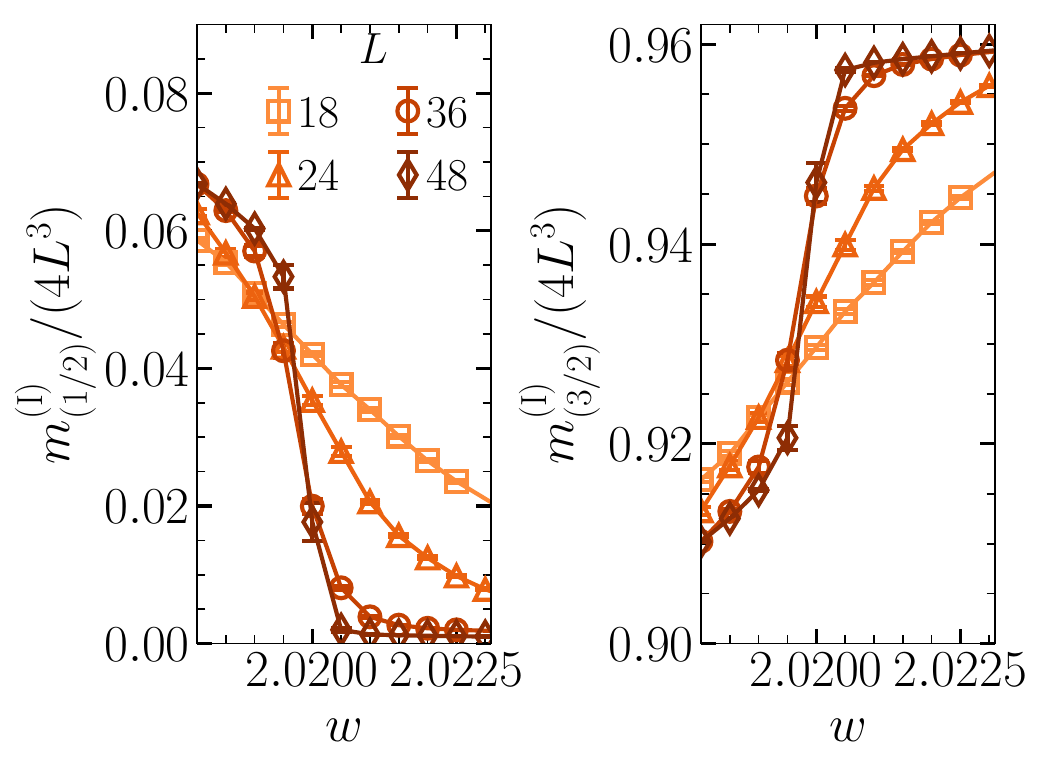}
    \caption{(a) The probability $P_{\rm cross}^{(1/2)}$ of having a connected cluster of spins with $S^z=\pm 1/2$
    shows clear evidence of a percolation transition: it tends to zero at large $L$ for $w>w_c$, while being nonzero in this limit for $w<w_c$. The corresponding probability for connected clusters of spins with $S^z=\pm 3/2$ is featureless and such clusters play no role in this transition. (b) The ratio of the masses of the second largest and largest connected clusters of spins with $S^z = \pm 1/2$ also shows a clear indication of this percolation transition, tending to zero at large $L$ for $w<w_c$ while remaining nonzero in this limit for $w>w_c$. Again, the corresponding quantity for $S^z = \pm 3/2$ clusters shows no signs of this transition. (c) The mass of the largest connected cluster of spins with $S^z = \pm 1/2$ scales with $L^3$ for $w<w_c$, but remains $O(1)$ in the large $L$ limit for $w>w_c$, consistent with the other indications of a percolation transition. The corresponding quantity for spins with $S^z = \pm 3/2$ scales with $L^3$ on both sides of $w_c$. }
    \label{fig:Pcross}
\end{figure*}
\begin{figure*}[h]
    \includegraphics[width=0.3\linewidth]{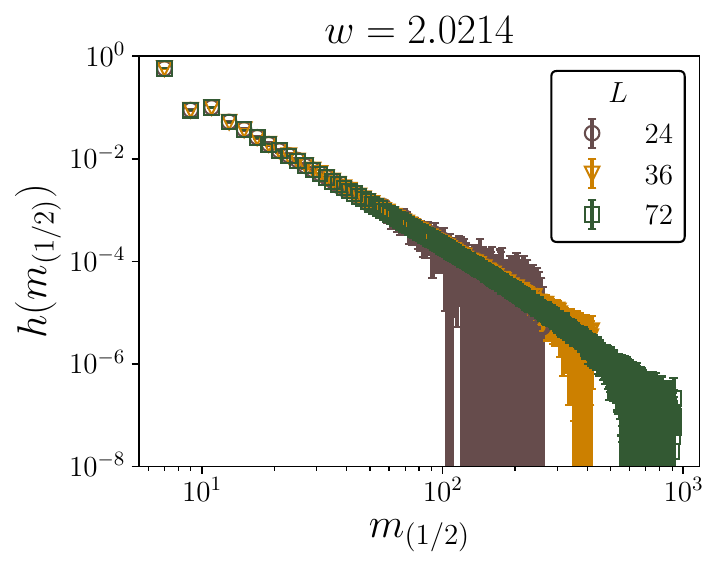}
    \includegraphics[width=0.3\linewidth]{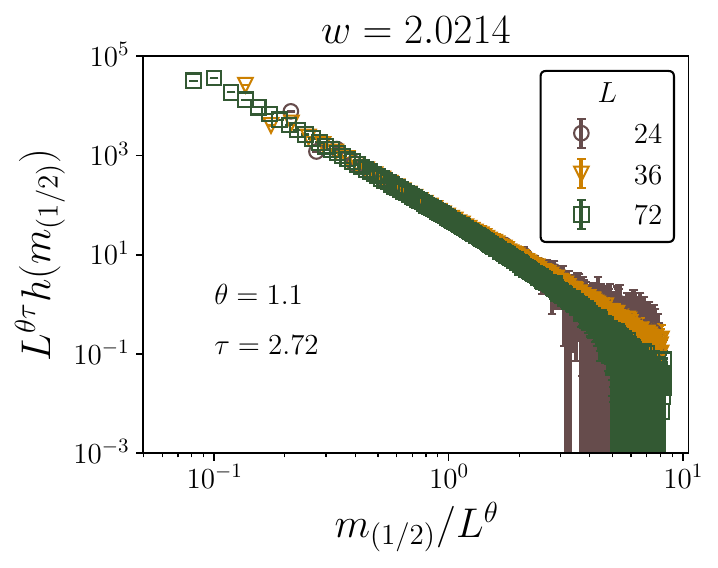}
     \includegraphics[width=0.3\linewidth]{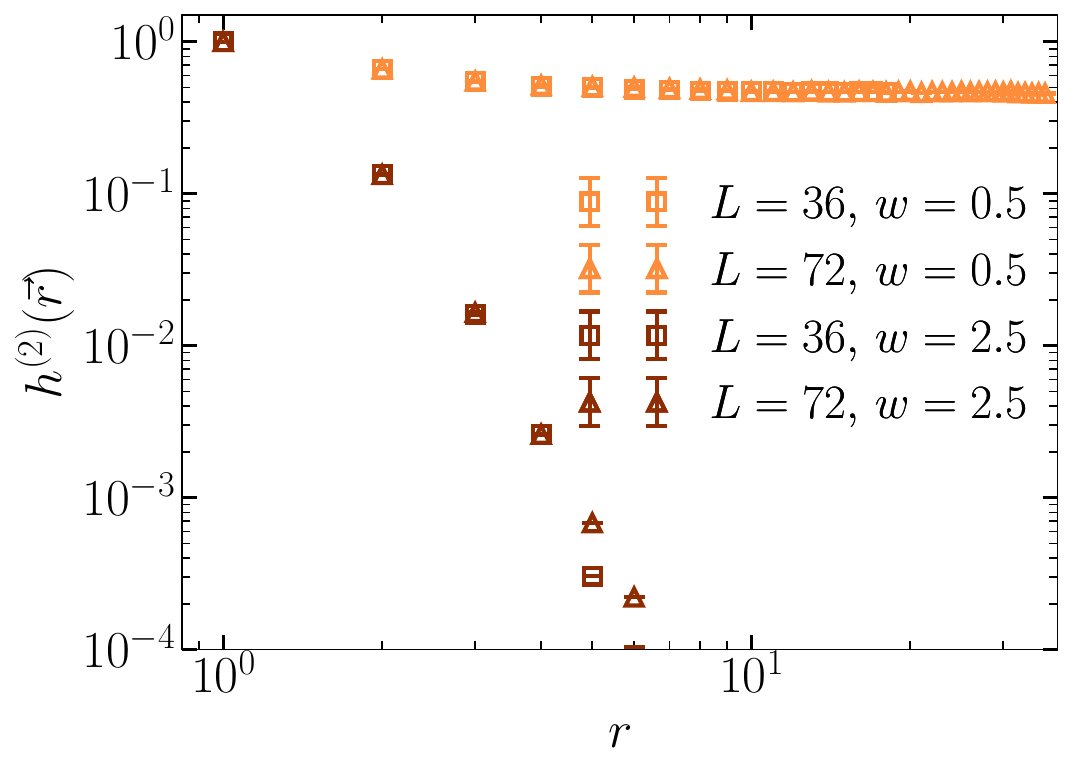}
    \caption{(a) For the sizes accessible to our numerics, the distribution of masses of clusters of spins with $S^z = \pm 1/2$ approximately follows a power-law form at criticality. (b) Indeed, this distribution approximately obeys the finite-size scaling form discussed in the Appendix. (c) The histogram $h^{(2)}(\vec{r})$ of head-to-tail displacements of charge-$2$ worms decays rapidly to zero at large $r$ for $w>w_c$. While this does not establish confinement of the corresponding test charge correlator $C^{(2)}(\vec{r})$, it is consistent with this.}
    \label{fig:ClusterHistogram}
\end{figure*}
The fraction of spins that have $S^z = \pm 3/2$ is quite high in the vicinity of $w_{c}$, with only about ten percent of the spins having $S^z = \pm 1/2$. What then drives the abrupt first-order transition to a $Z_3$ flux confined Coulomb liquid when $w$ increases beyond $w_c$?
The answer turns out to have a geometric flavor to it. We find that connected clusters of spins in the $S^z = \pm 1/2$ state undergo a weakly first-order percolation transition in the vicinity of $w_c$.

Indeed, we see multiple signatures of this geometric transition. First, we find that the probability $P_{\rm cross}^{(1/2)}$ that a single $S^z=\pm 1/2$ cluster winds around the three-dimensional torus in all three directions vanishes at large $L$ for $w>w_c$ but is nonzero in the thermodynamic limit for $w<w_c$ (Fig.~\ref{fig:Pcross}(a)). Thus $S^z=\pm 1/2$ clusters percolate in the small-$w$ phase but not the large-$w$ phase, and the transition at $w_c$ can be thought of as a percolation transition. However, when $P_{\rm cross}^{(1/2)}(w)$ is plotted against $w$ for fixed $L$, curves corresponding to different $L$ do not all cross at a single value of $w$, as expected at a continuous percolation transition. This $L$-dependent drift in the crossing point is consistent with behavior expected at a weakly first-order transition. 
The same behavior is also seen in the dimensionless ratio of the mass $m^{(II)}_{(1/2)}$ of the second-largest cluster of $S^z = \pm 1/2$ spins to the mass $m^{(I)}_{(1/2)}$ of the largest such cluster: This ratio goes to zero
 at large $L$ in the small-$w$ phase, but tends to a nonzero number in the large-$w$ phase. However, curves corresponding to different $L$ again do not all cross at a single value of $w$, but instead display a small drift in the crossing point as $L$ is increased (Fig.~\ref{fig:Pcross}(b) left panel). This is again consistent with a weakly first-order transition. Also, $m^{(I)}_{(1/2)}$ scales as $L^3$ in the small-$w$ phase, but not in
  the large-$w$ phase (Fig.~\ref{fig:Pcross}(c) left panel). 
In contrast, connected clusters formed by spins with $S^z = \pm 3/2$ percolate on both sides of $w_c$, and the largest such cluster remains macroscopic in size on both sides of $w_c$, as is clear from the data for the corresponding quantities for these connected clusters (inset of Fig.~\ref{fig:Pcross}(a) and right panels of Fig.~\ref{fig:Pcross} (b) (c)).

 Finally, the histogram $h(m_{(1/2)})$ of the masses of $S^z = \pm 1/2$ clusters at $w_c$ has an approximately power-law form (Fig.~\ref{fig:ClusterHistogram}(a)). Indeed, at the sizes readily accessible to our numerics, we find $h(m_{(1/2)})$ obeys a finite-size scaling form $h(m) \sim L^{-\theta \tau} f(m/L^{\theta})$ with $\theta \approx 1.1$ and $\tau \approx 2.72$ (Fig.~\ref{fig:ClusterHistogram}(b)). Thus, at the sizes accessible to us, this histogram displays behavior that is characteristic of a second-order critical point rather than a first-order transition. We therefore conclude that this $Z_3$ flux fractionalization transition between the two topologically distinct Coulomb liquids is driven by an interesting geometric percolation transition of clusters of $S^z = \pm 1/2$ spins. While the transition is clearly of a weakly first-order type, some finite size properties mimic behavior expected at critical points. 

Another interesting aspect of this $Z_3$ confinement transition has to do with the correlation function of test charges of charge $\pm 2$. As mentioned in the main text, partial information about this is provided by the histogram $h^{(2)}(\vec{r})$ of charge $\pm 2$ worms, in the sense that deconfined behavior of $h^{(2)}(\vec{r})$ implies the same for $C^{(2)}(\vec{r})$ but confinement seen in $h^{(2)}(\vec{r})$ does not immediately imply confinement for $C^{(2)}(\vec{r})$. Our picture in terms of $Z_3$ flux confinement suggests that $C^{(2)}(\vec{r})$ is confined for $w>w_c$. While we cannot unequivocally confirm this, we note that our results for $h^{(2)}(\vec{r})$ do show confinement, which is consistent with this expectation (Fig.~\ref{fig:ClusterHistogram}(c)).

\end{document}